# Wide Tuning Range and Low Noise Voltage Control Oscillators for 5G Technology

Authors: Minh Xuan Bui[1], Nguyen Thien Dat[1], Van Hong Lam[1], Tran Le Anh Quan[1], Pham Hung Anh[1], Mai Dong Xuan[2], Ke Wang[3]

Affiliation:

[1] School of Science Engineering and Technology, RMIT University Ho Chi Minh, Viet Nam

[2] Viettel Semiconductor Co., Ho Chi Minh, Viet Nam

[3] School of Engineering, RMIT University, Melbourne, Australia

## I. INTRODUCTION

The voltage control oscillator (VCO) is a key block of the phase lock loop, which generates the local oscillation signal for the transmitter and receiver of the wireless communication network. So far, there have been various approaches to design this analog intellectual property (IP) block. The circuit topology of the VCO can be classified as ring oscillator and LC tank-based oscillator. The ring oscillator combines a series of inverter blocks to form a 360-degree phase shift between the noise input and the voltage output of the oscillator, which acts as the positive feedback [1]. The main advantage of a ring oscillator is the simplicity of the circuit topology and the physical circuit layout design; however, tuning the oscillation frequency and sustaining stable oscillations remains a challenging barrier in the application of this method [2]. In contrast, an LC tank-based oscillator utilizes an inductor and capacitor tank to generate the resonant oscillation [3]. The amplifier using a field effect transistor (FET) compensates the LC tanks' power loss to maintain the oscillation condition at the output terminal of the VCO [4]. The LC VCO offers several key advantages over the ring oscillator, including superior frequency stability, as the LC tank circuit is less sensitive to temperature and supply voltage variations. It also delivers lower phase noise, crucial for high-frequency applications requiring signal purity, such as 5G communication systems [5]. Additionally, LC VCO provides a broader frequency range, making them suitable for both low and high-frequency applications [6], and they offer more linear frequency tuning, enabling precise control over oscillation frequency. These advantages make LC VCOs more reliable and efficient for advanced communication transceivers.

This paper presents the design anlysis of a new cascode cross-couple VCO, compares the performance with the conventional cross-couple one. Key design considerations are analyzed, including oscillation conditions, frequency range, and phase noise effects, highlighting the trade-offs and performance differences between these topologies. Examining these factors aims to provide valuable insights into the optimal VCO design choices for next-generation wireless communication systems.

## II. ANALYTICAL DESIGN OF VCO

To improve the tuning range, gain and phase noise, a new cascode cross-coupled LC VCO was proposed as shown in figure 1. This new topology adds another cross-coupled pair below the conventional pair with an inductor connected between the oscillator and ground. Regarding DC bias operation, the drain and gate voltages of M1 and M3 are equal, thus they are operating in saturation mode. On the other hand, the drain voltage of M2 and M4 are lower than gate voltage of M2 and M4, hence these two transistors operate in the triode mode. The small-signal model of the proposed cascode cross-couple pair is shown in figure 2. Since M1 and M3 are matched and operate in saturation mode, their small signal models can be presented by output resistance between drain and source ($r_{o1} = r_{o3}$), transconductance ($g_{m1} = g_{m3}$) and the dominant impedance between gate and source ($Z_{gs1} = Z_{gs3}$). Since M2 and M4 are matched and operate in the triode mode, their small signal models can be represented as resistance ($\frac{1}{g_{m2}}$) between drain and source and dominant impedance between gate and source ($Z_{gs2}$). The same analysis procedure as in section $A$ was followed to determine its start-up oscillation condition and the oscillation frequency.

- *Start-up oscillation condition*

The start-up condition is determined based on the real part of the input admittance and the equivalent resistance of the LC tank:

$$\frac{4}{g_m} = 2r_{o1} \parallel R_P \quad (1)$$

- *Oscillation frequency:*

$$f_o(cascode) = \frac{1}{2\pi\sqrt{L(1.75C_{gs} + C_P + C_L + C_{var})}} \quad (2)$$

### A. Phase noise anlysis

Signal integrity is crucial in the VCO design process, especially for 5G applications. Noise reduction methodologies and techniques must be carried out to minimize signal interference and provide a pure output signal. The main factors contributing to the phase noise of the VCO are thermal and flicker noise, which are generated by passive and active devices, respectively, and noise from the cross-coupled pair. The single sideband noise spectral density of the VCO can be estimated with the following equation [7]:

$$L(\Delta\omega) = 10\log_{10}\left[\frac{1}{2}\frac{KT}{V_{max}^2}\frac{1}{R_p(C_{total}\omega_o)^2}\left(\frac{\omega_o}{\Delta\omega}\right)^2\right] \quad (3)$$

where $\omega_0$ is the oscillation frequency, $\Delta\omega$ is the offset frequency from the carrier, $K$ is the Boltzman constant, $T$ is the absolute temperature, $R_p$ is the equivalent parallel resistor, $C_{total}$ is the total equivalent capacitance of the LC tank, and $V_{max}$ is the maximum voltage swing across the tank.

As shown in figure 3, the total equivalent capacitance of the LC tank of the proposed cascode VCO:

$$C_{total}(cascode) = 0.875C_{gs} + \frac{C_p + C_L + C_{var}}{2} \quad (4)$$

Equation (4) shows that the proposed cascode topology results in an increase of the total equivalent capacitance of LC tanks, thus leading to the reduction of the phase noise spectrum density of the VCO as evident in equation (3). Furthermore, equation (1) shows that the proposed cascode topology leads to a double increase of the negative resistance, compared to the conventional VCO. This indicates the increase of the equivalent parasitic resistance of the LC tanks ($R_p$), which can be derived from equation (1). As a result, the phase noise spectrum density of the VCO is reduced thanks to the increase of the equivalent parasitic resistance of the LC tank.

## III. Simulation Results and Discussion

### B. Tuning Range and VCO Gain

The tuning ranges of the proposed cascode cross-couple and the conventional cross-couple VCOs are compared in figure 4. It is clear that the tuning range of the proposed VCO is 5GHz (from 21.0 to 26.1 GHz), which is significantly higher than 4.2GHz of the conventional VCO (from 22.6 to 26.8 GHz). It can also be seen from the simulation results in figure 5 that the VCO gain (slope of the graph) at the middle point of the tuning range (Vtune=0.5V) of the proposal VCO (8GHz/V) is much higher than that of the conventional one (5.3GHz/V).

### C. Phase Noise

The phase noise comparison between the proposed and conventional VCOs is presented in figure 6. With the offset frequency from 10 MHz to 800 MHz, the phase noise spectrum of the proposed VCO is consistently lower than that of the conventional one. Specifically, at the frequency of 800 MHz, the proposed VCO results in the phase noise spectrum of -155.7 dBc/Hz, which is lower than -154.4 dBc/Hz of the conventional VCO.

### D. Discussion

Simulation results and theoretical analysis demonstrate that the proposed VCO topology outperforms the conventional, most visible through oscillation frequency tuning range, VCO gain and phase noise. The broader tuning range and higher gain of the proposed cascode cross-coupled oscillator can be explained as the consequence of the increased total parasitic capacitance of the VCO as shown in equation (4). Furthermore, the improvement of the phase noise of the proposed cascode VCO is also attributed to the increase of the equivalent parallel resistance ($R_p$) of the LC tank thanks to the addition of the negative resistance as shown in equation (1).

## IV. Conclusion

The paper compared the performance of the proposed cascode cross-couple and the conventional cross-couple LC voltage control oscillators (VCO) in terms of turning range, gain and phase noise. Small signal model of the VCO and admittance analysis were presented to evaluate the start-up oscillation condition, oscillation frequency and the phase noise effect of two VCO topologies. Theoretical analysis and simulation results show that the proposed cascode cross-couple VCO results in higher tuning range, higher VCO gain and lower phase noise compared with the conventional cross-couple VCO thanks to the increase of the total equivalent parasitic capacitance and the equivalent parasitic resistance of the LC tank.

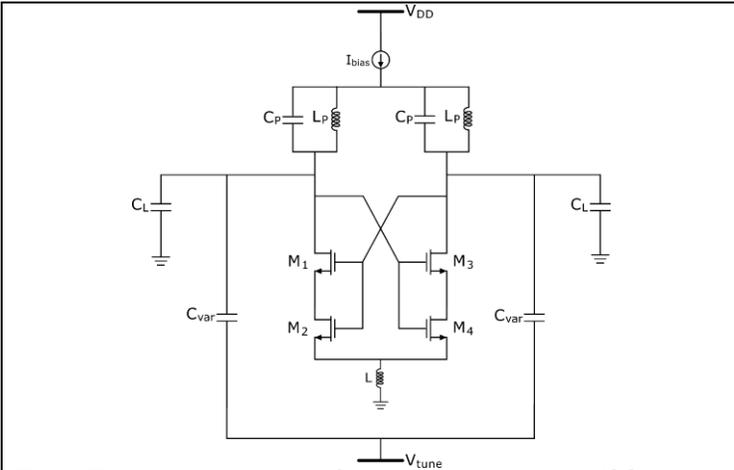

Fig. 1: The schematic circuit of the proposed cascode VCO

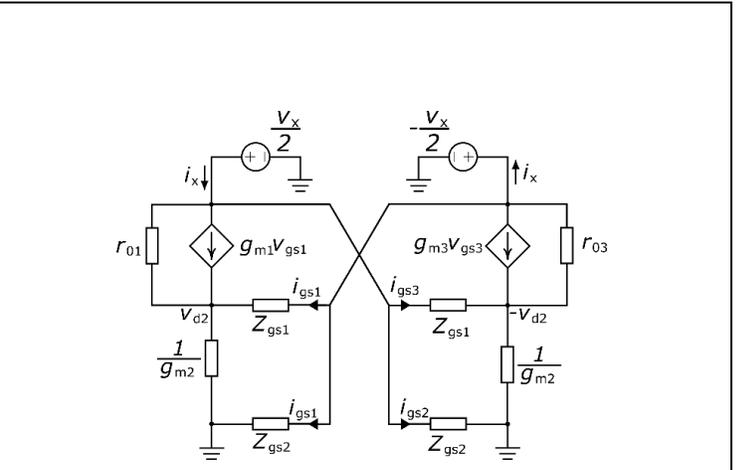

Fig. 2: The small-signal model of proposed cascode pair

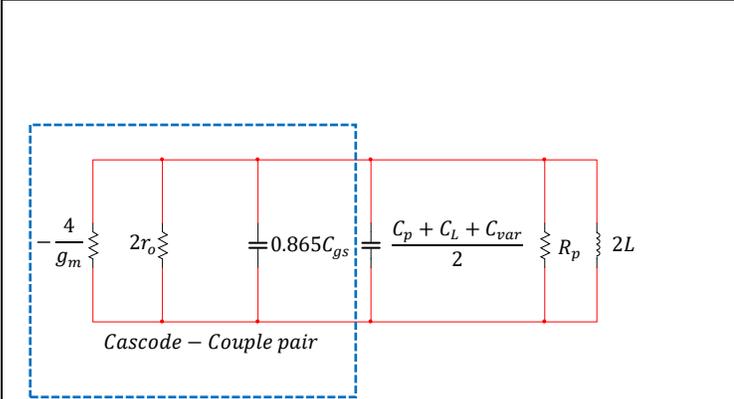

Fig. 3: Equivalent small signal model of the proposed cascode VCO

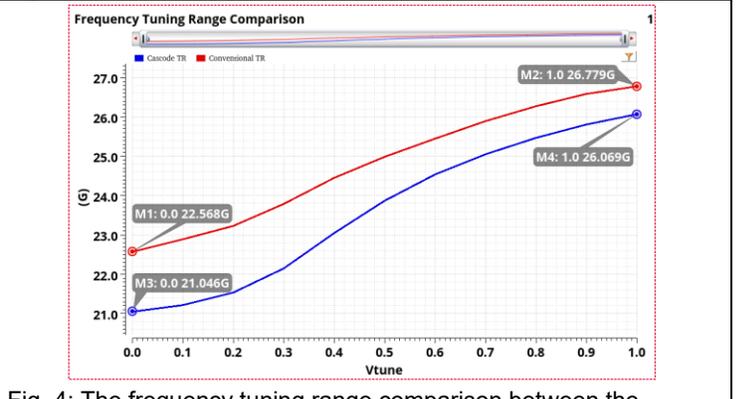

Fig. 4: The frequency tuning range comparison between the proposed and conventional VCOs

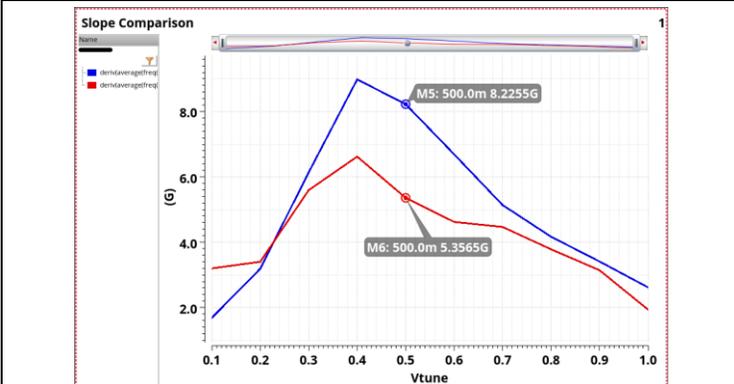

Fig. 5: The VCO gain (slope) comparison between the proposed and conventional VCOs

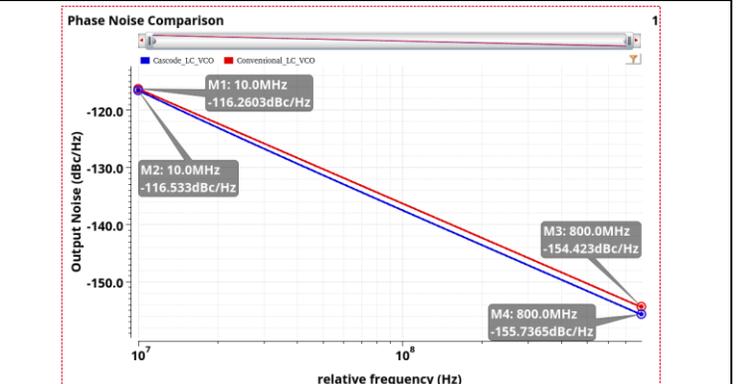

Fig. 6: The phase noise comparison between the proposed and conventional VCOs